\documentclass[%
 reprint,
 superscriptaddress,
 amsmath,amssymb,
 prb,
 nofootinbib,
]{revtex4-2}

\usepackage[usenames,dvipsnames]{xcolor}
\usepackage{amsthm}
\usepackage{amssymb}

\usepackage{diagbox}

\usepackage{graphicx}% Include figure files
\usepackage{dcolumn}% Align table columns on decimal point
\usepackage{bm}% bold math
\usepackage[unicode, colorlinks, urlcolor=blue, linkcolor=blue, pagecolor=blue, citecolor=blue]{hyperref} 

\usepackage[english]{babel}

\usepackage{comment}
\addto\captionsenglish{}
\usepackage{bigstrut}
\usepackage{float}

\let\oldcfrac\cfrac
\renewcommand{\cfrac}[2]{\oldcfrac{#1}{#2}\,}

\usepackage{comment}
\usepackage{hhline}
\usepackage{ stmaryrd }
\usepackage{snapshot}

\begin{document}

\title{Mechanical properties of V-4Ti-4Cr alloy from molecular dynamics with a neural-network potential}

\newcommand\JIHT{Joint Institute for High Temperatures, Izhorskaya 13 Bldg 2, Moscow 125412, Russia}
\newcommand\MIPT{Moscow Institute of Physics and Technology, Institutskiy Pereulok 9, Dolgoprudny, Moscow Region, 141701, Russia}
\newcommand\JINR{Joint Institute for Nuclear Research, Zholio-Kyuri 6, Dubna, 141980, Russia}
\newcommand\HPPI{Institute for High Pressure Physics, 108840, Moscow, Troitsk,  Russia}

\author{G. S. Demyanov}
\affiliation{\JIHT}
\affiliation{\MIPT}
\author{D. V. Minakov}
\affiliation{\JIHT}
\affiliation{\MIPT}
\author{S.B. Saltykov}
\affiliation{\HPPI}
\author{P. R. Levashov}
\affiliation{\JIHT}
\affiliation{\MIPT}
\author{N. M. Chtchelkatchev}
\affiliation{\JIHT}
\affiliation{\JINR}

\date{\today}

\begin{abstract}
Machine-learning interatomic potentials enable large-scale atomistic simulations of vanadium alloys relevant to fusion applications, but reliable training and validation remain challenging in multicomponent systems. Here, we develop a descriptor-based DeepMD-DPA1 potential for the V--Ti--Cr system using a two-stage workflow: broad configuration sampling driven by the MatterSim foundation model followed by fine-tuning to density-functional-theory data computed with VASP. The model achieves root-mean-square errors of 9.2~meV/atom for energies and 0.23~eV/\AA{} for force components. Using this potential in large-cell LAMMPS simulations, we compute Young's modulus, bulk modulus, and Poisson's ratio for V--4Ti--$x$Cr and V--$x$Ti--4Cr alloys at $T=300$~K and $T=1073$~K. We find that increasing Cr fraction increases the elastic moduli of the alloy, while increasing Ti fraction decreases them; all compositions are softer at $1073$~K than at $300$~K. Two-phase simulations give a melting temperature for pure V of 1950~K, in good agreement with experimental data.
\end{abstract}

\maketitle

\section{Introduction}
Vanadium-based alloys are promising structural materials for fusion reactor first-wall and blanket components due to their low induced radioactivity, good high-temperature mechanical performance, and compatibility with liquid lithium systems \cite{Chung1995FED,Muroga2014JNM}. Among them, V--4Ti--4Cr is a reference low-activation composition \cite{Muroga2014JNM}.

The elastic response of V--4Ti--4Cr has been determined in resonance ultrasound experiments at $\sim$100~kHz over 100--300~K. In particular, the room-temperature Young's modulus shows a noticeable dependence on the specific material heat-treatment and processing history: for two independently produced V--4Ti--4Cr heats, the reported values are $E=127$~GPa and $E=121$~GPa, respectively (for comparison, a closely related V--5Ti--5Cr heat gives $E=124$~GPa) \cite{KardashevChernov2008FTT}.
Hydrogen charging further modifies the elastic response: it reduces $E$ relative to the hydrogen-free state and produces a pronounced minimum in $E(T)$ near 180~K, accompanied by an internal-friction peak around 140~K \cite{Kardashev2022}. The melting temperature of pure vanadium is measured experimentally using pulse-heating and containerless techniques~\cite{Pottlacher2007_liquidV,Paradis2002_ESL_V,McClure1997_radianceV}.

%The mechanical properties and microstructure of this alloy have been studied experimentally under different processing routes and environments \cite{Muroga2014JNM,Potapenko2015}. For example, hydrogen can measurably affect Young's modulus and internal friction in V--4Ti--4Cr \cite{Kardashev2022}. The melting temperature of pure vanadium is also apparent experimentally from pulse-heating and containerless measurements, and it is often used as a reference point in high-temperature thermophysical work and temperature-scale studies \cite{Pottlacher2007_liquidV,Paradis2002_ESL_V,McClure1997_radianceV,Bedford1984_SRP,Pavese1996_ITS90_SRP}.

However, the performance of V--4Cr--4Ti is sensitive to interstitial impurities (O, N, and C) and gas species (hydrogen isotopes and He), all of which affect embrittlement, precipitation, and the irradiation response \cite{Muroga2014JNM,Watanabe2007JNM}. Titanium can strengthen the alloy and getter interstitials into Ti(C, O, N)-type precipitates \cite{Sparks2022JNM}. Here, we focus on the substitutional V--Ti--Cr solid alloy. Explicit treatment of interstitial solutes and precipitation is beyond the scope of this study.

Atomistic simulations can address temperature and composition dependence of mechanical properties by resolving defect energetics, diffusion, and thermomechanical response. However, direct density functional theory (DFT) calculations are too expensive for large cells and long trajectories \cite{Muroga2014JNM}. In particular, \textit{ab initio} molecular dynamics (MD) is usually limited to hundreds of atoms and short time scales, which is not sufficient for melting or stress--strain testing~\cite{Wang2018DeePMDKit,Zhang2020DPGEN}. 

Machine-learning interatomic potentials (MLIPs) provide a practical route to large-scale simulations by learning the DFT potential energy surface and enabling molecular dynamics (MD) at near classical MD cost while retaining near-DFT accuracy for energies, forces, and virials \cite{Wang2018DeePMDKit,Zhang2020DPGEN,Behler2021,Jacobs2025,Zhang2025}. Building reliable MLIPs typically requires systematic sampling and dataset refinement, including active-learning loops that add new DFT labels where needed \cite{Zhang2020DPGEN}. We use a descriptor-based DeepMD model, which is computationally efficient and has a mature training and fine-tuning workflow \cite{Wang2018DeePMDKit}.

In this work, we develop and validate a DeepMD-DPA1 potential for the V--4Ti--4Cr alloy~\cite{Potapenko2015}. We use a two-stage dataset strategy: (i) generate a large set of configurations by classical MD driven by the MatterSim foundation model~\cite{yang2024mattersim} across random V--Ti--Cr compositions and temperatures, and (ii) fine-tune the model on a subset recomputed with the Vienna \textit{Ab initio} Simulation Package (VASP) \cite{Kresse:PRB:1996}. We validate the potential using two-phase melting and stress--strain simulations at ambient pressure to estimate the melting temperature and elastic properties~\cite{Sparks2022JNM,Kardashev2022}. Also, we vary the Ti and Cr concentrations by assigning atomic types in a fixed supercell. This makes it straightforward to scan many V-$x$Ti-$y$Cr compositions with the same simulation technique and to identify trends that can guide experimental studies.

The paper is organized as follows. Section~\ref{sec:mlip} describes the dataset generation and the DeepMD-DPA1 training workflow. Section~\ref{sec:mech} details the molecular-dynamics algorithms used to compute elastic properties. The two-phase method to estimate the melting temperature is summarized in Section~\ref{sec:Tm}. The main results for melting and composition- and temperature-dependent elastic properties are presented in Section~\ref{sec:res}. Conclusions and limitations are discussed in Section~\ref{sec:conc}.

\section{Deep learning potential and training procedure \label{sec:mlip}}
In this work, we use a single machine-learning interatomic potential trained in the DeepMD framework~\cite{Wang2018DeePMDKit}. The final model is based on the DeepMD-DPA1 architecture and is constructed in two stages: (i) pretraining on a large, foundation-model-generated dataset to obtain broad coverage of the relevant configuration space, followed by (ii) fine-tuning on a smaller set of DFT calculations performed by VASP to restore DFT-level accuracy for energies, forces, and stresses.

\textit{Foundation-model sampling.} As a first step, we generated an extensive pool of V--Ti--Cr configurations using MD driven by the MatterSim foundation model~\cite{yang2024mattersim}. The trajectories covered random alloy compositions and temperatures from 300~K to 2000~K. We ran MD in both the NPT ensemble at near-ambient pressure and the NVT ensemble. In the NVT runs, we applied a random perturbation of the cell volume by up to 10\% every 1000 MD steps to accelerate exploration of strained and non-equilibrium environments. Sampling included large supercells ($\sim 500$ atoms) as well as smaller cells of roughly 60--100 atoms. The MD time step was 1~fs, and each trajectory was 10--20~ps long. To further promote chemical disorder and diversify local environments, we randomly swapped the atomic identities among all species every 1000 MD steps.

\begin{figure}[t]
\centering
\includegraphics[width=\linewidth]{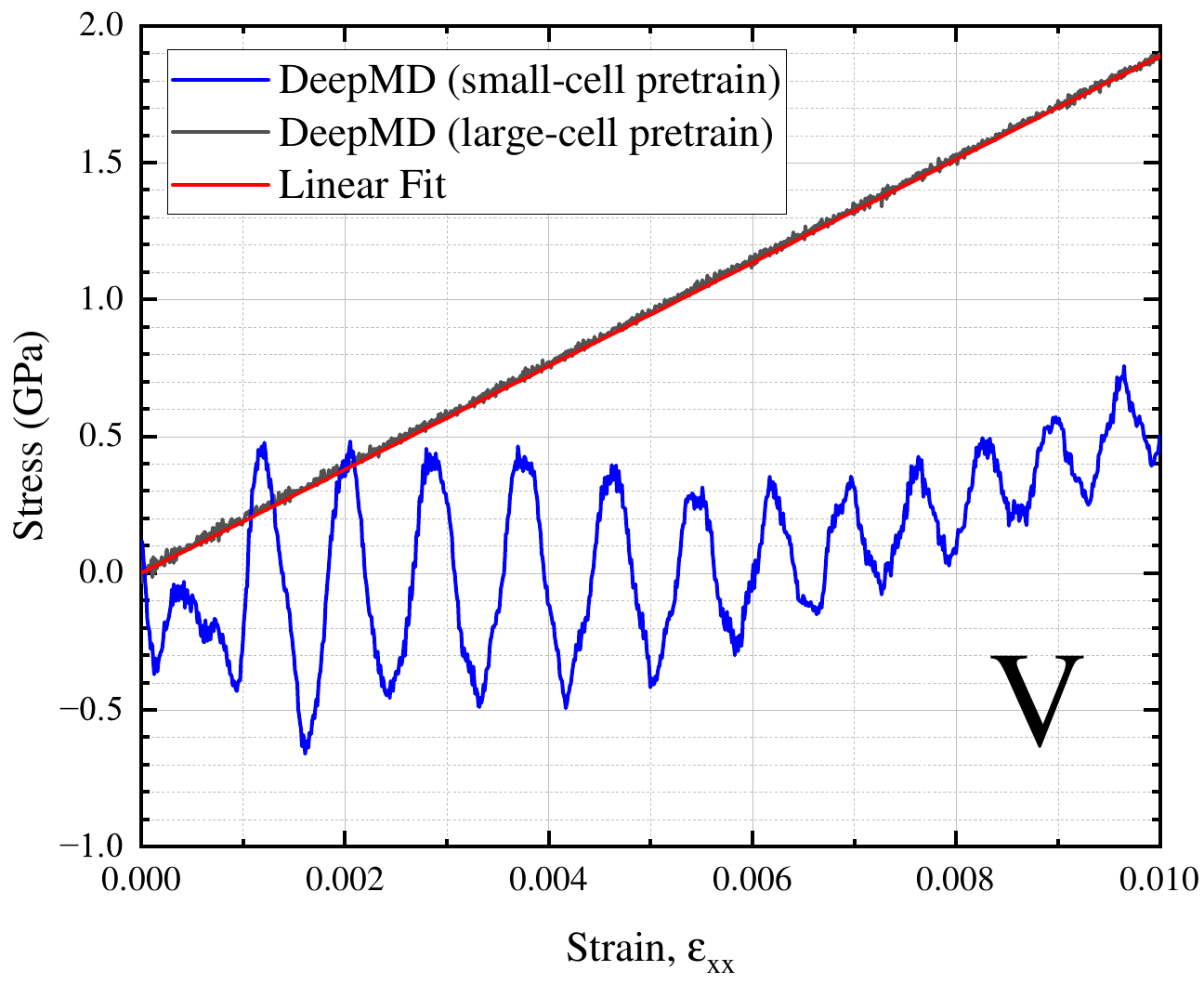}
\caption{Axial stress--strain response of pure bcc V at $T~=~300$~K comparing two DeepMD models that differ only in the configuration sizes used at the MatterSim pretraining stage. When pretraining is performed on small configurations (up to $\sim 100$ atoms), the axial stress $\sigma_{xx}$ exhibits pronounced oscillations (including negative values) during tensile loading. Expanding the pretraining dataset with additional large-cell trajectories ($\sim 500$ atoms) eliminates these oscillations and yields a smooth elastic response. The red line indicates the linear fit in the small-strain regime used to extract Young's modulus.}
\label{fig:stress_size_artifact}
\end{figure}

We note that the cell sizes present at the pretraining stage can matter for mechanical validation. In particular, a preliminary model pretrained only on small foundation-model configurations (up to $\sim 100$ atoms) and then fine-tuned on the same VASP subset exhibited pronounced oscillations of the axial stress (including sign changes) in large-cell stress-strain tests (see Fig.~\ref{fig:stress_size_artifact}). To mitigate this artifact, the final pretraining dataset included additional \texttt{large\_cell} trajectories with $\sim 500$ atoms; after this update, the tensile stress--strain response became stable without non-physical oscillations, and this final model was used throughout the present work.

\textit{Selection and DFT labeling.} From the foundation-model trajectories, we selected a subset of configurations for DFT labeling. For the small-cell NPT trajectories, we randomly subsampled 200--300 structures per compound to reduce redundancy while preserving diversity. The resulting DFT subset contained 1400 configurations and was recomputed with VASP to obtain reference energies, forces, and stresses. We used the Perdew-Burke-Ernzerhof (PBE) parametrization
for the generalized gradient approximation (GGA)~\cite{PhysRevLett.77.3865} as an exchange--correlation (XC) functional. Five electrons
were treated as valence in the framework of the projector augmented-wave (PAW) pseudopotential~\cite{PhysRevB.59.1758}.

\textit{Training and fine-tuning.} We first train a DeepMD-DPA1 potential on the large MatterSim dataset (several hundred thousand configurations), which provides a robust initialization and allows efficient training on broad configuration coverage. We then fine-tune this model on the DFT-PBE dataset. During both dataset preparation and training, we filter rare configurations exhibiting extremely short interatomic distances and very large forces, because such configurations can destabilize optimization and degrade the resulting model quality.

\begin{figure*}[t]
\centering
\begin{minipage}[t]{0.32\textwidth}
\centering
\includegraphics[width=\linewidth]{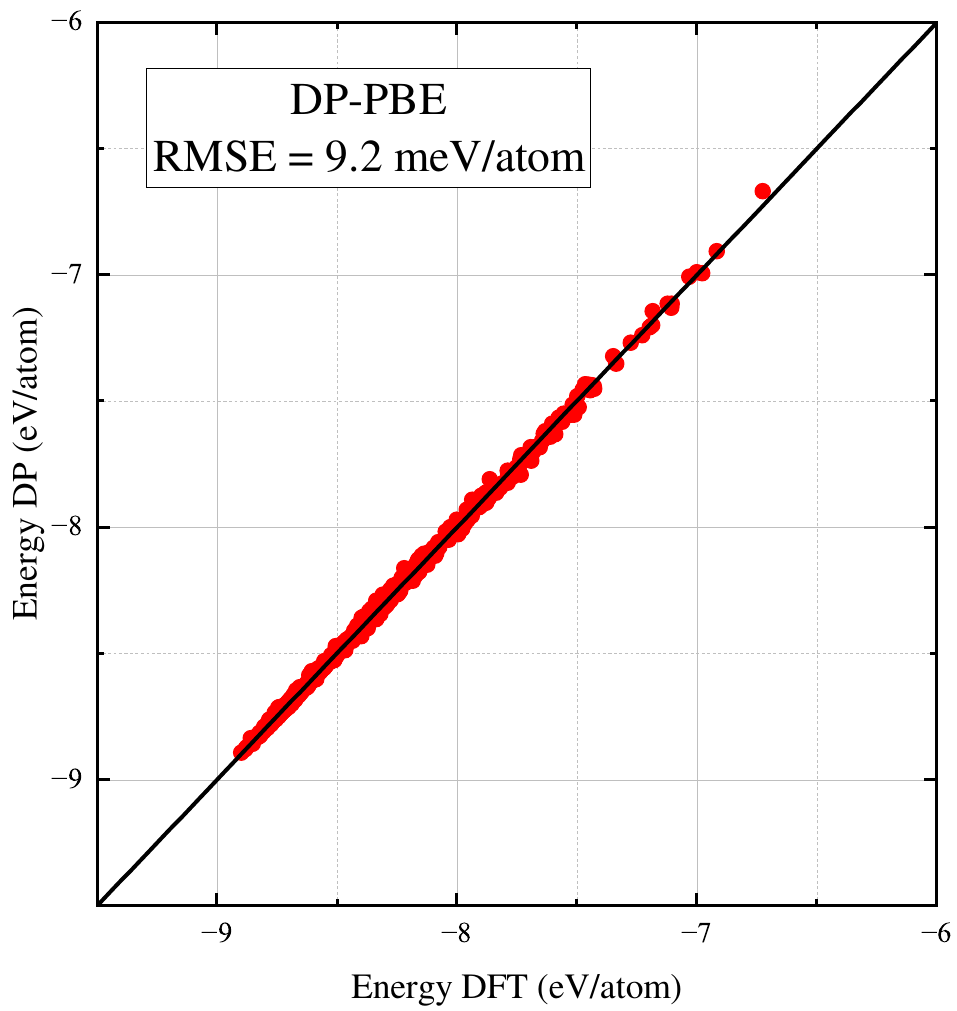}
\par\small\textbf{(a)} Energy per atom
\end{minipage}
\hfill
\begin{minipage}[t]{0.32\textwidth}
\centering
\includegraphics[width=\linewidth]{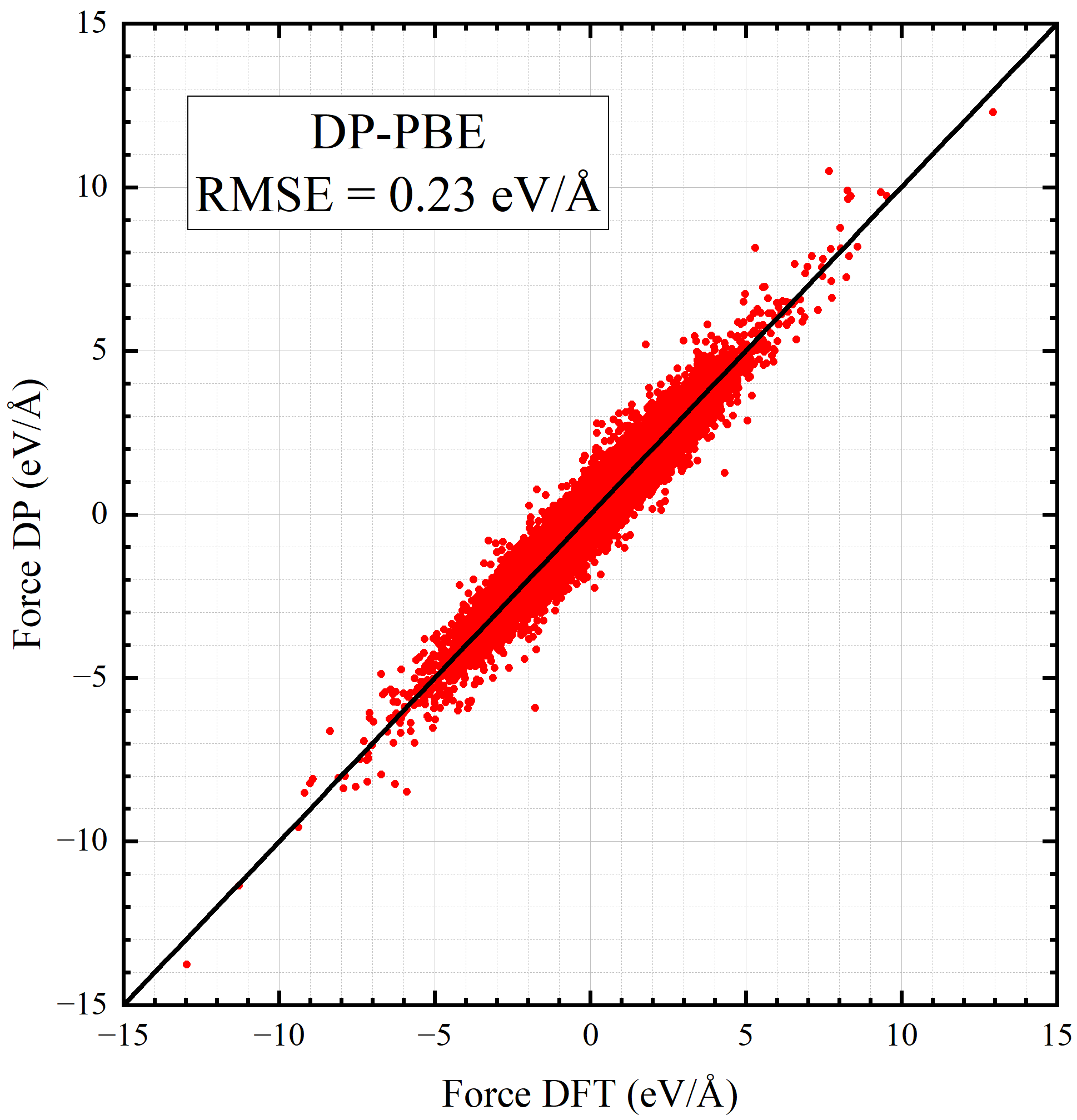}
\par\small\textbf{(b)} Force
\end{minipage}
\hfill
\begin{minipage}[t]{0.32\textwidth}
\centering
\includegraphics[width=\linewidth]{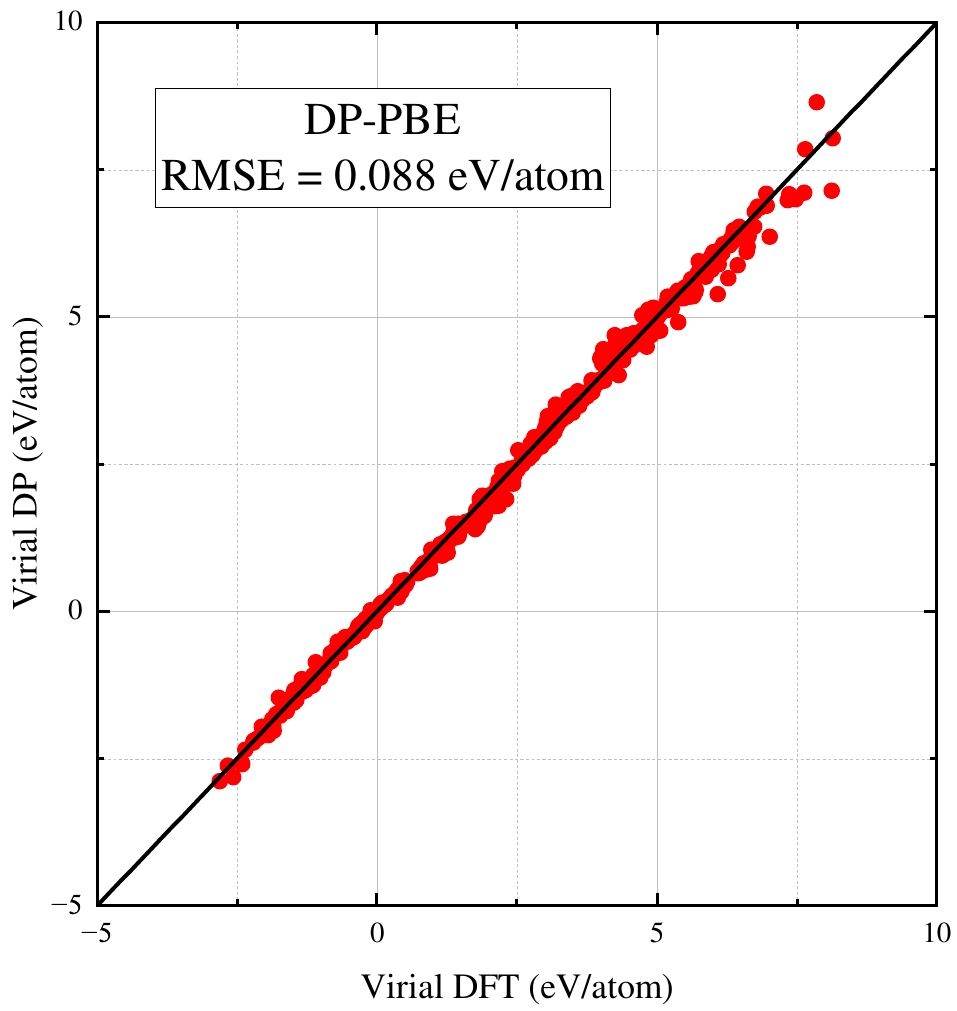}
\par\small\textbf{(c)} Virial per atom
\end{minipage}
\caption{Parity plots comparing DeepMD predictions to DFT reference values on the evaluation set. Panels show (a) energies per atom, (b) x-force component, and (c) virial per atom. The black diagonal line indicates perfect agreement.}
\label{fig:rmse_parity}
\end{figure*}

\textit{Model accuracy.} The trained potential was evaluated against the reference DFT data. The resulting mean absolute errors (MAE) and root-mean-square errors (RMSE) are summarized in Table~\ref{tab:mlip_accuracy}, and representative parity plots are shown in Fig.~\ref{fig:rmse_parity}.

\begin{table}[h!]
\caption{\label{tab:mlip_accuracy}Accuracy metrics of the trained DeepMD-DPA1 potential on the evaluation set.}
\begin{ruledtabular}
\begin{tabular}{lcc}
Quantity & MAE & RMSE \\
\hline
Energy (meV/atom) & 5.6 & 9.2 \\
Forces (eV/\AA) & 0.17 & 0.23 \\
Stress (meV/\AA$^{3}$) & 2.9 & 6.2 \\
Virial (meV/atom) & 37 & 71 \\
\end{tabular}
\end{ruledtabular}
\end{table}

\section{Method for calculating mechanical properties \label{sec:mech}}
Mechanical properties were evaluated by MD simulations in LAMMPS \cite{Thompson:CPC:2022} using the trained interatomic potential. We performed uniaxial tensile loading to extract Young's modulus and Poisson's ratio, isotropic compression to obtain the bulk modulus, and simple shear to obtain the shear modulus. In the current setups, the uniaxial simulations used a $30\times 30\times 30$ bcc supercell ($N=54000$ atoms), while the bulk and shear simulations used $20\times 20\times 20$ bcc supercells ($N=16000$ atoms). We considered variable alloy compositions, denoted as VTi$_x$Cr$_y$, where $x$ and $y$ are the Ti and Cr atomic concentrations (in at.\%); the compositions were set by assigning atomic types in the initial supercell. All mechanical-property calculations were performed at two temperatures, $T=300$~K and $T=1073$~K.

We first equilibrated the system in the NPT ensemble at the target temperature and zero pressure, which defined the reference cell parameters (cell lengths $L_{x,0}$, $L_{y,0}$, $L_{z,0}$, and volume $V_0$). The MD time step was $\Delta t=1$~fs. Temperature and pressure were controlled with a Nos\'e--Hoover thermostat and barostat. 

\begin{figure}[t]
\centering
\includegraphics[width=0.95\linewidth]{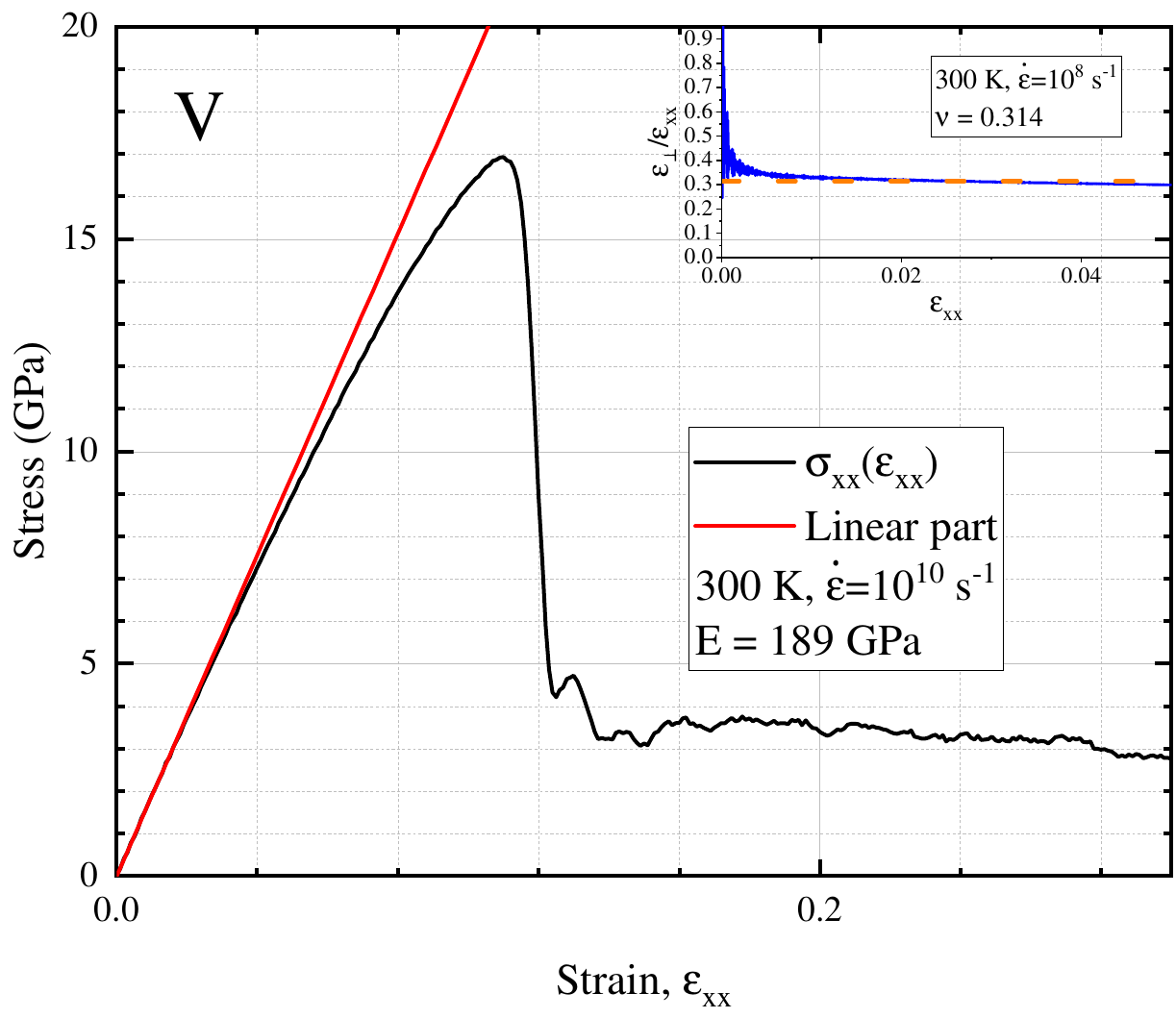}
\caption{Representative axial stress--strain curve used to extract Young's modulus from the initial linear regime. The plot shows $\sigma_{xx}$ as a function of $\varepsilon_{xx}$ for pure V at $T=300$~K (strain rate as indicated in the legend). The red line highlights the small-strain interval selected for a least-squares fit; beyond this window, the curve departs from linear elasticity and exhibits a pronounced stress drop, so it is excluded from the modulus evaluation. The inset shows the corresponding Poisson's ratio $\nu=-\varepsilon_{\perp}/\varepsilon_{xx}$ computed during loading and illustrates the averaging procedure used to report $\nu$ in the elastic regime.}
\label{fig:stress_strain_example}
\end{figure}

To calculate Young's modulus and Poisson's ratio, we applied a uniaxial deformation along the $x$ direction while maintaining zero pressure in the transverse directions with an NPT barostat~\cite{Maulana2020AIP}. The axial strain is computed from the instantaneous box length $L_x$ as
\begin{equation}
\varepsilon_{xx}=\frac{L_x-L_{x,0}}{L_{x,0}},
\end{equation}
where $L_{x,0}$ is the equilibrated reference length. The transverse strain is computed from the average transverse length $L_d=(L_y+L_z)/2$:
\begin{equation}
\varepsilon_{\perp}=\frac{L_d-L_{d,0}}{L_{d,0}}. %, \qquad L_d=\frac{L_y+L_z}{2}.
\end{equation}
LAMMPS reports the pressure tensor components $p_{\alpha\beta}$, so the stress tensor can be calculated directly as $\sigma_{\alpha\beta}=-p_{\alpha\beta}$. Young's modulus is obtained from the linear relation between the axial stress and strain,
\begin{equation}
E=\left.\frac{d\sigma_{xx}}{d\varepsilon_{xx}}\right|_{\varepsilon_{xx}\to 0},
\end{equation}
evaluated by a linear fit of $\sigma_{xx}(\varepsilon_{xx})$ in the elastic regime (see Fig.~\ref{fig:stress_strain_example}). Poisson's ratio is computed from its definition and averaged over a selected time window (or, equivalently, over a range of strains within the elastic regime):
\begin{equation}
\nu=-\left\langle \frac{\varepsilon_{\perp}}{\varepsilon_{xx}} \right\rangle.
\end{equation}
In practice, we compute an instantaneous ratio $\nu(t)=-\varepsilon_{\perp}(t)/\varepsilon_{xx}(t)$ from the box dimensions at every output step and then average it over the same elastic window used for the linear fit of $E$.
An example of the resulting $\nu(\varepsilon_{xx})$ trace is shown in the inset of Fig.~\ref{fig:stress_strain_example}.

The bulk modulus is computed from an isotropic deformation in which the box is compressed at equal rates in all three directions. We define the volumetric strain from the instantaneous volume $V$ as
\begin{equation}
\varepsilon_V=\frac{V-V_0}{V_0},
\end{equation}
where $V_0$ is the equilibrated reference volume, and we use the mean pressure $p=(p_{xx}+p_{yy}+p_{zz})/3$. In the linear regime, isotropic elasticity gives $p=-K\,\varepsilon_V$ (positive $p$ for compression, negative $\varepsilon_V$ for a volume decrease), so the bulk modulus is obtained from the slope
\begin{equation}
K=-\left.\frac{dp}{d\varepsilon_V}\right|_{\varepsilon_V\to 0},
\end{equation}
using a linear fit similar to the Poisson's ratio calculation.

The shear modulus is obtained from a simple shear deformation applied to a triclinic cell, where the $xz$ tilt factor is varied at a constant rate. The shear strain is computed as
\begin{equation}
\gamma_{xz}=\frac{xz}{L_z},
\end{equation}
and the corresponding shear stress is $\tau_{xz}=\sigma_{xz}=-p_{xz}$. In the elastic regime, $\tau_{xz}=G\,\gamma_{xz}$, and the shear modulus is obtained from
\begin{equation}
G=\left.\frac{d\tau_{xz}}{d\gamma_{xz}}\right|_{\gamma_{xz}\to 0}.
\end{equation}
For an isotropic solid, these moduli are related by
\begin{equation}
\label{eq:isotropic_relations}
E=\frac{9KG}{3K+G}, \qquad \nu=\frac{3K-2G}{2(3K+G)}.
\end{equation}
%We use these relations as an internal consistency check when applicable; any deviations indicate either nonlinearity/plasticity in the sampled regime or residual finite-temperature noise, and therefore motivate restricting the fits to smaller strains or longer averaging windows.

Stress and strain were recorded during the loading, and elastic constants were obtained from the linear slope of the stress--strain curves in the small-strain regime, with averaging over multiple steps to reduce noise. Deformation was applied at constant strain rates. We used $\dot{\varepsilon}$ in the $10^8$--$10^{10}$~s$^{-1}$ range (Fig.~\ref{fig:stress_strain_example}), while the value for isotropic compression and shear was $10^8$~s$^{-1}$. For all moduli, we fitted the slope over a small elastic window $\varepsilon\in[0,\varepsilon_\mathrm{max}]$, with $\varepsilon_\mathrm{max}$ chosen to avoid nonlinear or plastic response.

\section{Method for calculating the melting temperature \label{sec:Tm}}
The melting temperature was determined using the standard two-phase method. We constructed a periodic slab cell containing coexisting solid and liquid, with the interface normal to the $z$ axis. The cell was initialized as bcc V with a lattice parameter $a=3.024$~\AA{} and dimensions $15\times 15\times 45$ unit cells, and we split the box into two equal halves along $z$ to define the initial solid and liquid regions. For a bcc lattice (2 atoms per unit cell), this corresponds to $N=20250$ atoms in total.

To prepare a stable two-phase starting point, we equilibrated the two halves separately: the liquid region was held at a high temperature $T_\mathrm{melt}$ where the system is fully liquid, and the solid region at a lower temperature $T_\mathrm{solid}$ where the crystal remains stable. After this pre-equilibration, the whole system was assigned a common temperature $T_\mathrm{phase}$ and evolved in the NPT ensemble at the target external pressure (zero). The MD time step was $\Delta t=1$~fs.

The melting temperature $T_m$ is defined as the temperature at which the solid--liquid interface shows no systematic drift over the averaging time window. For temperatures $T_\mathrm{phase} > T_m$, the solid region melts and the liquid phase grows, whereas for $T_\mathrm{phase} < T_m$, the liquid crystallizes and the solid region expands.
In practice, the instantaneous position of the solid--liquid interface $z_\mathrm{int}(t)$ is extracted from the atomic trajectories, and the interface velocity is defined as
\begin{equation}
v_\mathrm{int}(T_\mathrm{phase}) = \left\langle \frac{d z_\mathrm{int}}{dt} \right\rangle,
\end{equation}
where $\langle \cdot \rangle$ denotes a time average performed over a steady-state interval after discarding the initial transient.
The melting temperature $T_m$ is determined from the condition
\begin{equation}
v_\mathrm{int}(T_m) = 0.
\end{equation}
In practice, $T_m$ is obtained by performing simulations at several trial temperatures and identifying the temperature interval in which the sign of $v_\mathrm{int}$ changes. Linear interpolation of $v_\mathrm{int}(T)$ between the two nearest temperatures with opposite signs is used to estimate $T_m$. The width of this bracketing interval provides a practical estimate of the uncertainty arising from finite simulation time and finite system size.

\section{Results and discussion \label{sec:res}}

\begin{figure*}[t]
\centering
\includegraphics[width=0.49\textwidth]{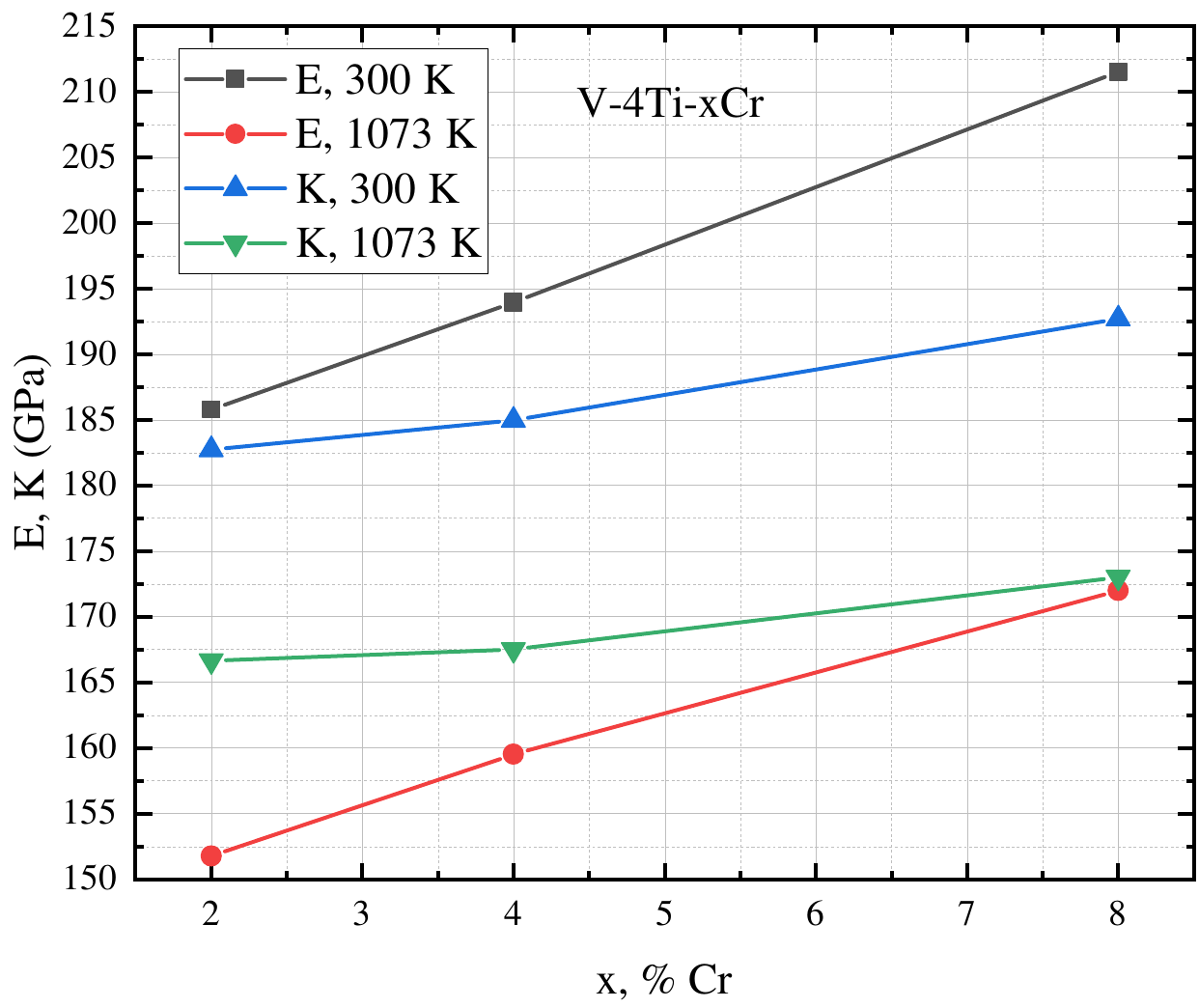}
\includegraphics[width=0.49\textwidth]{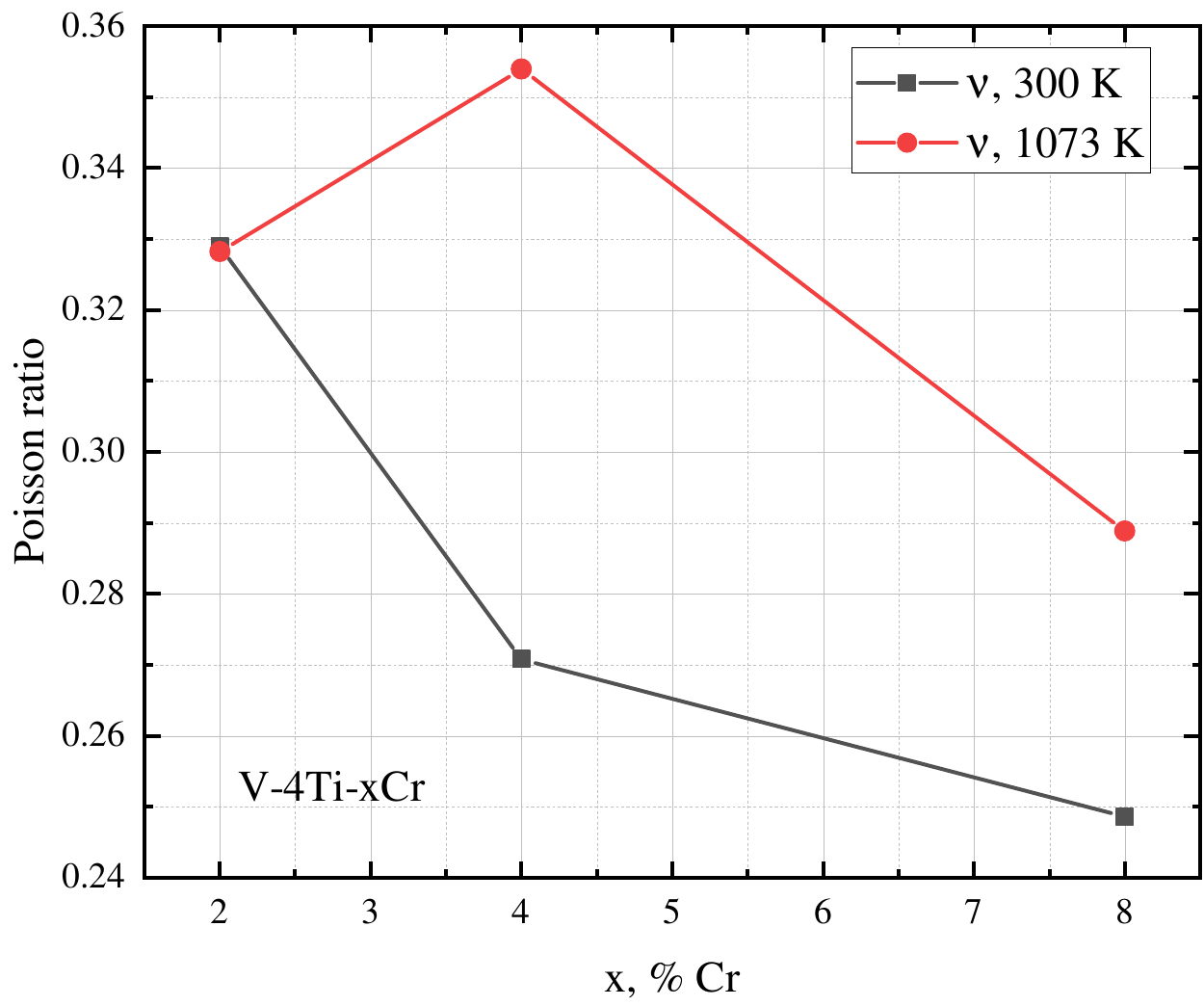}
\caption{Elastic properties of V--4Ti--$x$Cr as a function of Cr concentration $x$ at $T=300$~K and $T=1073$~K. Left: Young's modulus $E$ and bulk modulus $K$. Right: Poisson's ratio $\nu$.}
\label{fig:elastic_v4ti_xcr}
\end{figure*}

This section summarizes the composition and temperature dependence of the elastic response predicted by our MD simulations for V--Ti--Cr alloys. We report Young's modulus $E$, bulk modulus $K$, and Poisson's ratio $\nu$ for two alloy families: V--4Ti--$x$Cr (variable Cr content at fixed Ti) and V--$x$Ti--4Cr (variable Ti content at fixed Cr). For each composition, the moduli are calculated at $T=300$~K and $T=1073$~K using the procedures described above.

\begin{figure*}[t]
\centering
\includegraphics[width=0.49\textwidth]{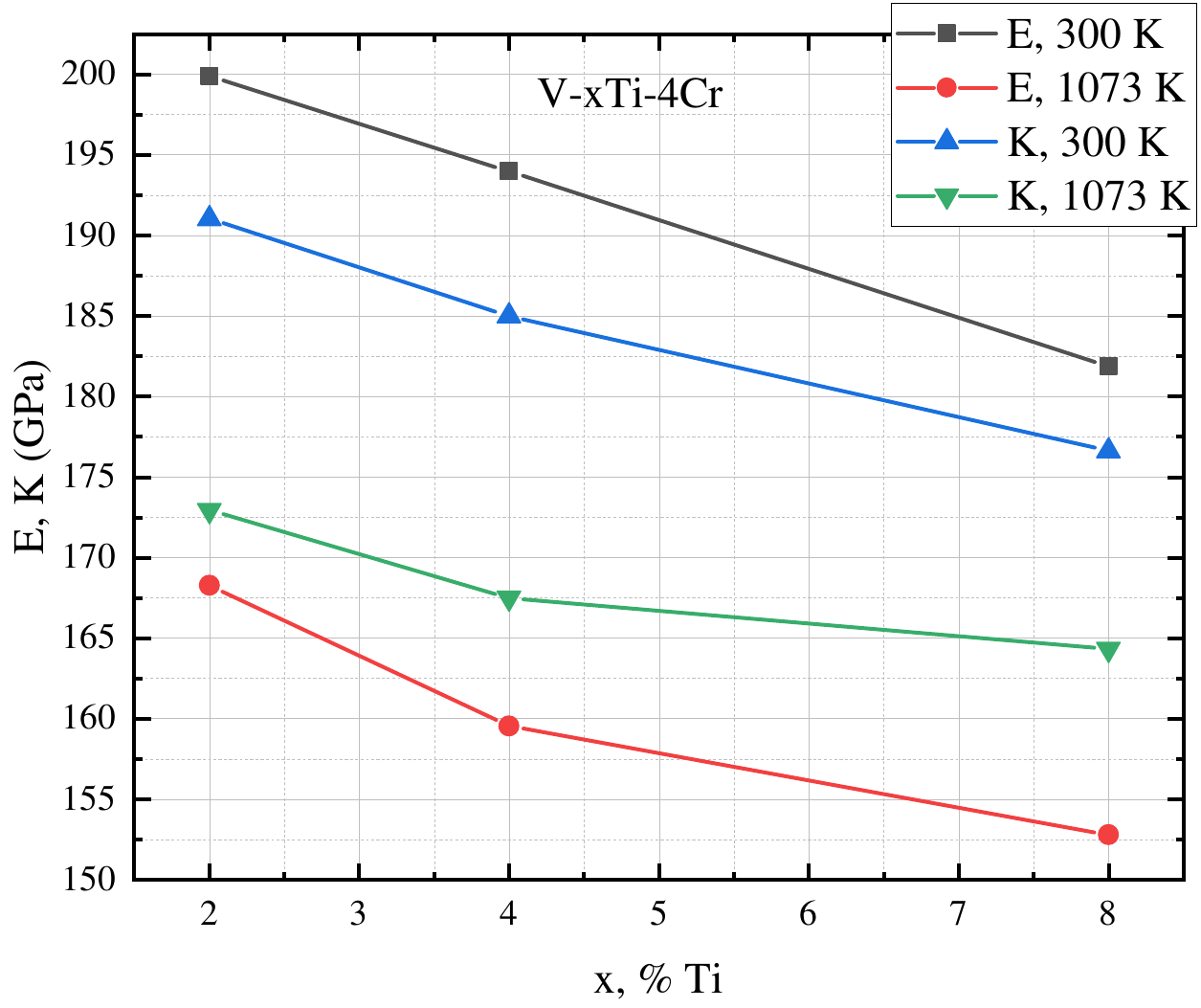}
\includegraphics[width=0.49\textwidth]{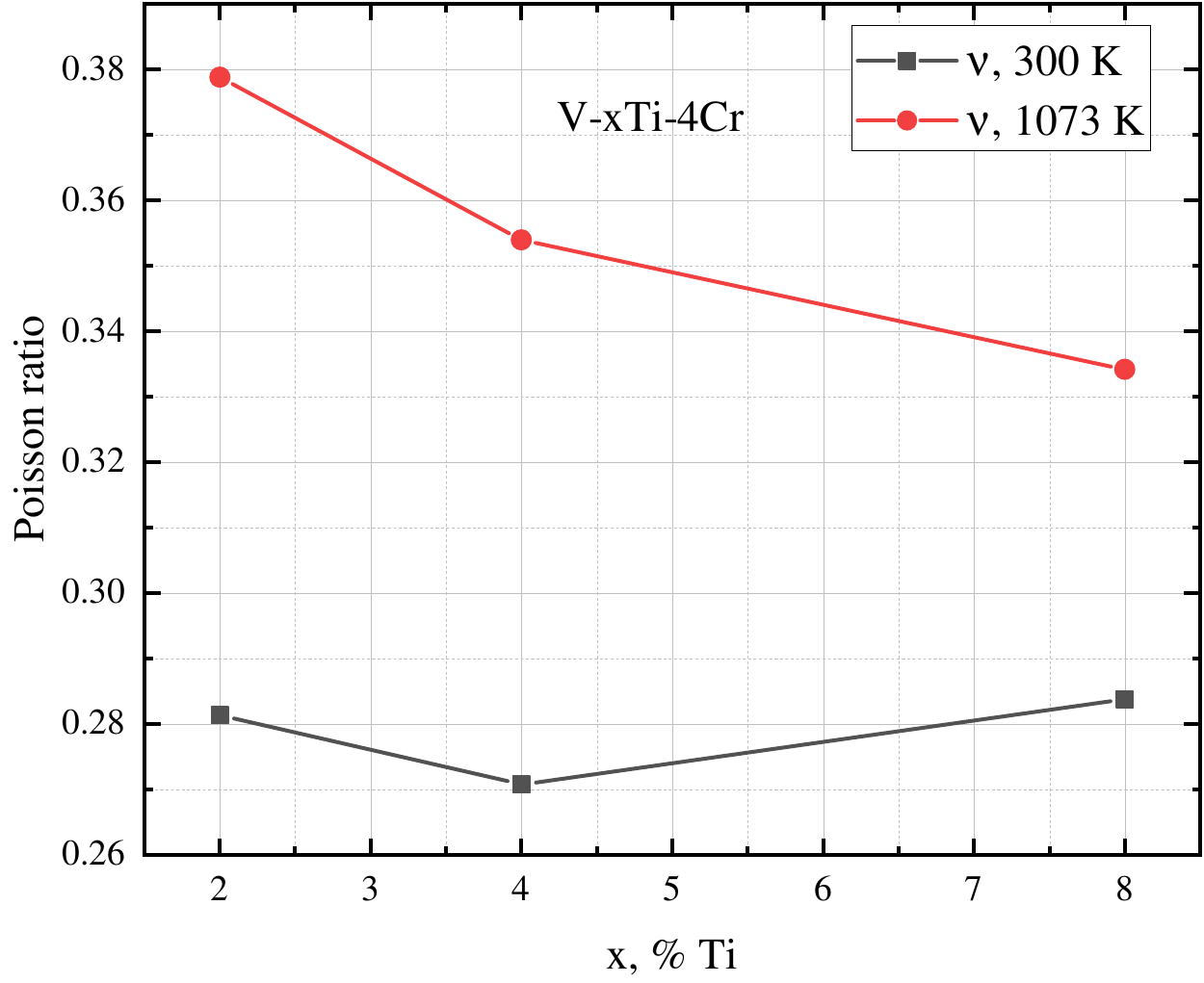}
\caption{Elastic properties of V--$x$Ti--4Cr as a function of Ti concentration $x$ at $T=300$~K and $T=1073$~K. Left: Young's modulus $E$ and bulk modulus $K$. Right: Poisson's ratio $\nu$.}
\label{fig:elastic_vx_ti4cr}
\end{figure*}

Before describing results, we report two validation observations obtained for pure bcc V in our stress-strain test. First, we found that the pretraining dataset can strongly affect mechanical stability. In early versions of our workflow, a DeepMD-DPA1 model pretrained only on a small number of MatterSim configurations (up to $\sim 100$ atoms) produced pronounced, wave-like oscillations of the axial stress during tensile loading at $T=300$~K, including transient sign changes (Fig.~\ref{fig:stress_size_artifact}). Expanding the MatterSim pretraining dataset with additional large-cell trajectories ($\sim 500$ atoms) eliminated these oscillations (see Fig.~\ref{fig:stress_size_artifact}). The VASP fine-tuning dataset was identical in both cases, so the effect is attributed to the pretraining procedure rather than to the DFT labeling set. 

Second, we observed that the absolute magnitude of Young's modulus is sensitive to whether the model is fine-tuned to DFT. The pretraining-only model (without VASP fine-tuning) yields $E\approx 140$~GPa for pure V at $T=300$~K, close to typical experimental values of $\sim 130$~GPa~\cite{met15040427, KardashevChernov2008FTT}.  After DFT-PBE fine-tuning, $E$ increases to $180$~GPa, while preserving the expected thermal softening ($E$ decreases with $T$). Thus fine-tuning drives the potential toward the DFT elastic response. This interpretation is consistent with our comparison to PBE elastic constants available from Ref.~\cite{Elastic_Principles_2014}. Therefore, discrepancies with experiment are likely due to the choice of exchange--correlation functional and the use of ideal, defect-free reference structures, rather than a fundamental limitation of the MLIP approach.

We also evaluated the melting temperature of pure vanadium using the two-phase protocol described in Sec.~\ref{sec:Tm}. The estimated melting temperature is \mbox{$1968\pm25$~K.} For comparison, experimental measurements report a melting temperature of about 2199~K \cite{Pottlacher2007_liquidV}. Given that the training dataset includes configurations only up to $2000$~K, the predicted melting temperature shows reasonable agreement with experimental data.

For V--4Ti--$x$Cr, both $E$ and $K$ increase with increasing Cr concentration within the considered range (see Fig.~\ref{fig:elastic_v4ti_xcr}). This indicates that the alloy stiffens due to changes in composition. The variation of $K$ with composition is weaker than that of $E$. Poisson's ratio exhibits more complex behavior. At $300$~K, it decreases as the Cr fraction increases. At $1073$~K, it shows non-monotonic dependence with a maximum at an intermediate Cr concentration. At the higher temperature, the alloy is systematically softer (lower $E$ and $K$) than at $300$~K, as expected.

For V--$x$Ti--4Cr, the predicted $E$ and $K$ decrease with increasing Ti concentration (Fig.~\ref{fig:elastic_vx_ti4cr}), i.e., Ti additions soften the alloy within the sampled compositions. In contrast to the clear trends in $E$ and $K$, the Poisson's ratio is weakly composition-dependent at $300$~K, whereas at $1073$~K it decreases with increasing Ti fraction. Across both alloy compounds, $\nu$ is higher at $1073$~K than at $300$~K, while $E$ and $K$ are lower, which is consistent with the expected increase of elastic compliance at elevated temperature.

Finally, we emphasize that interpreting absolute elastic constants requires care. For isotropic elasticity, Eq.~(\ref{eq:isotropic_relations}) shows that $E$ is governed primarily by the shear response $G$; shear moduli are particularly sensitive to microstructural defects (dislocations, vacancies, impurities) that are absent in ideal DFT cells and in our defect-free reference crystals. In preliminary tests, introducing point defects reduces $E$, which qualitatively supports this interpretation. Future work will therefore assess the dependence of elastic constants on the exchange--correlation functional used for fine-tuning (e.g., PBE vs SCAN) and incorporate representative defect populations when comparing to experiments.

\section{\label{sec:conc}Conclusion}
We developed and validated a DeepMD-DPA1 potential for the V--Ti--Cr alloy using MatterSim foundation-model sampling followed by VASP-PBE fine-tuning. The model enables large-cell finite-temperature simulations and systematic compositional screening of V-$x$Ti-$y$Cr alloys while consistently simulating mechanical properties.

For the two alloy compounds studied, increasing the Cr fraction at fixed Ti increases Young's and bulk moduli, whereas increasing the Ti fraction at fixed Cr decreases them. All compositions show thermal softening as temperature increases from $T=300$~K to $T=1073$~K. The Poisson's ratio varies more weakly with composition and is higher at the elevated temperature. Using the two-phase method, we estimate the melting temperature of pure V as $1968$~K, below the experimental value of 2199~K \cite{Pottlacher2007_liquidV}, which points to remaining sensitivity to the underlying DFT reference and training coverage at high temperature.

We also find that the pretraining dataset should include large supercell configurations in order to avoid spurious stress oscillations in tensile simulations. In future work, we plan to refine the DFT setup and investigate the choice of exchange--correlation functional, extend the melting calculations to alloy compositions, and examine the role of representative defects when comparing computed elastic constants with experimental data.

\begin{acknowledgments}
    The work is supported by the Russian Science Foundation (project No. 25-29-01649). The authors acknowledge the JIHT RAS Supercomputer Centre and the Shared Resource Centre ``Far Eastern Computing Resource'' IACP FEB RAS for providing computing time. Part of the computations were carried out on MVS-10P at the JointSupercomputer Center of the Russian Academy of Sciences (JSCCRAS) and using the computing resources of the Federal Collective Usage Center Complex for Sim-ulation and Data Processing for Mega-Science Facilities at NRC“Kurchatov Institute,” http://ckp.nrcki.ru/.  

\end{acknowledgments}

\appendix

\bibliography{refs}

\end{document}